\documentclass[conference]{IEEEtran}
\makeatletter
\def\ps@headings{%
\def\@oddhead{\mbox{}\scriptsize\rightmark \hfil \thepage}%
\def\@evenhead{\scriptsize\thepage \hfil \leftmark\mbox{}}%
\def\@oddfoot{}%
\def\@evenfoot{}}
\makeatother
\usepackage{tikz}
\usepackage{xurl}
\usepackage{float}
\usepackage{bbm}
\usepackage{amsmath}
\usepackage{graphicx}
\usepackage{textcomp}
\usepackage{xcolor}
\usepackage{gensymb}
\usepackage{xspace}
\usepackage{booktabs}
\usepackage{multirow}
\usepackage{balance}
\usepackage{pifont}
\usepackage{caption}
\usepackage{subcaption}
\usepackage[noend]{algpseudocode}
\usepackage[linesnumbered,ruled]{algorithm2e}
\usepackage[tmargin=0.75in, bmargin=1.1in, lmargin=0.63in, rmargin=0.635in]{geometry}

\def\eg{e.g.\xspace}
\def\S{Hippo\xspace}
\def\ie{\textit{i.e.}\xspace}

\newtheorem{thm}{Theorem}[section]
\newtheorem{lem}[thm]{Lemma}

\def\BibTeX{{\rm B\kern-.05em{\sc i\kern-.025em b}\kern-.08em
    T\kern-.1667em\lower.7ex\hbox{E}\kern-.125emX}}

\begin{document}

\title{Protecting Activity Sensing Data Privacy Using Hierarchical Information Dissociation}

\author{
\IEEEauthorblockN{
Guangjing Wang, Hanqing Guo, Yuanda Wang, Bocheng Chen, Ce Zhou, Qiben Yan
}
\IEEEauthorblockA{Department of Computer Science, Michigan State University, East Lansing, MI, USA}%

\IEEEauthorblockA{\{wanggu22, guohanqi, wangy208, chenboc1, zhouce, qyan\}@msu.edu}%
}

\maketitle

\begin{abstract}
Smartphones and wearable devices have been integrated into our daily lives, offering personalized services. However, many apps become overprivileged as their collected sensing data contains unnecessary sensitive information. For example, mobile sensing data could reveal private attributes (e.g., gender and age) and unintended sensitive features (e.g., hand gestures when entering passwords). To prevent sensitive information leakage, existing methods must obtain private labels and users need to specify privacy policies. However, they only achieve limited control over information disclosure. In this work, we present Hippo to dissociate hierarchical information including private metadata and multi-grained activity information from the sensing data. Hippo achieves fine-grained control over the disclosure of sensitive information without requiring private labels. Specifically, we design a latent guidance-based diffusion model, which generates multi-grained versions of raw sensor data conditioned on hierarchical latent activity features.  Hippo enables users to control the disclosure of sensitive information in sensing data, ensuring their privacy while preserving the necessary features to meet the utility requirements of applications. Hippo is the first unified model that achieves two goals: perturbing the sensitive attributes and controlling the disclosure of sensitive information in mobile sensing data.  Extensive experiments show that Hippo can anonymize personal attributes and transform activity information at various resolutions across different types of sensing data. 
\end{abstract}

\label{abstract}


\section{Introduction}
\label{sec:intro}
Smartphones and wearable devices are equipped with motion and position sensors that measure motion and orientation. The mobile devices provide the sensing data to various applications or third parties. Currently, mobile operating systems (OS) classify motion and position sensors as low-risk sensors, which means any app could access the raw sensing data without any permission restriction. For instance, Android allows an app to access sensor data at a sampling rate exceeding 200 Hz without explicit user consent~\cite{google2023permission}. Thus, a travel app that offers a compass bearing using motion sensors might access arbitrary activity-sensing data over an entire day. 

Generally, mobile sensing data contains not only the information required by applications but also redundant and sensitive information that users do not intend to share, resulting in concerns related to over-privilege and user privacy leakage~\cite{yeke2023wear, han2019shad}. There are two key overprivileged issues in mobile activity sensing data: (i) \textit{Metadata-level overprivileged issue}: it allows apps to pinpoint users' private attributes such as age and gender. Activity-sensing data contains redundant personal attribute information, which is unnecessary for activity-sensing functionalities but has privacy implications. Users with distinct attributes perform physical activities differently. In consequence, attackers could exploit these unique features within the sensor data to infer users' sensitive attributes~\cite{malekzadeh2019mobile}. 

(ii) \textit{Feature-level overprivileged issue}: it allows apps to collect more fine-grained activity features than necessary. The fine-grained activity features can potentially compromise users' sensitive behaviors~\cite{enck2014taintdroid}. For instance, an elderly person wears a smartwatch for fall detection. While the primary purpose is safety, the collected sensor data can also track precise hand movements~\cite{cheon2020gesture}. That means, by capturing the raw sensor data,  a malicious service provider could potentially steal passwords entered by the users for their bank accounts~\cite{shukla2019stealing}. Generally, malicious app developers can exploit the redundant fine-grained features in sensor data to learn sensitive behaviors and task-irrelevant concepts~\cite{song2020overlearning}. 
Therefore, what appears to be low-risk sensors may indeed carry a high privacy risk, given that fine-grained sensor data could compromise user privacy.

Many research efforts have been devoted to coping with cases where data required for specific tasks is  privacy-sensitive. In this work, we mainly consider the overprivileged issues that partial information in the data is task-unrelated but privacy-sensitive. For metadata-level overprivileged issues, adversarial training~\cite{liu2019privacy, boutet2021dysan} has been utilized to perturb personal attributes linked to sensor data, but they need to collect labeled private data for model training. For feature-level overprivileged issues, various filtering mechanisms~\cite{gotz2012maskit, chakraborty2014ipshield} have been proposed to limit data collection. For instance, mobile OS could control whether to release or hide an entire segment of raw data using privacy filtering mechanisms. However, existing solutions are impractical as they require specific privacy policies or labels of private data from users for data filtering. Moreover, current OS policy-based filtering methods are coarse-grained as they either wholly release or remove segments of raw data.  If a data segment corresponds to a private activity, the whole data segment would be discarded, thereby damaging the data utility. 

A naive solution is to add noise~\cite{he2018preserving}, which can potentially perturb both private attributes and activity features. However, the sensor data has recurring patterns in continuous temporal windows. To prevent data cleaning methods from filtering out the added noise, existing methods need to introduce a substantial amount of noise into each window~\cite{malekzadeh2018replacement}, which will greatly destroy the data utility (\ie, activity pattern recognition). As a result, existing noise-based methods are either ineffective or could destroy most of the data utility. 

To tackle the above problems, we propose \S to limit the information in data by reconstructing multi-grained mobile activity sensing data. By integrating the idea of autoencoders and generative models, we design the latent feature guidance-based diffusion model to dissociate hierarchical information. \S learns to extract different granularity of latent feature representations by learning concept hierarchy from the data. Through the hierarchical information learning process, \S dissociates the fine-grained information including sensitive attributes (metadata) and multi-resolution hierarchical activity features in the raw sensor data. \S allows the users to remove sensitive information that they do not intend to disclose, in order to protect sensitive activity-sensing data. \S can be used as a middleware between OS and applications for private attributes and sensitive feature dissociation in raw sensor data.

There are two main challenges in designing \S. \textit{First, how to perturb the metadata embedded in the sensor data while retaining raw activity features without labeling private data for training?} Adversarial training based on labeled private data has been widely applied to perturb private attributes. Each private attribute needs to be labeled if multiple attributes are to be perturbed. Such supervised training methods are impractical considering the difficulty of obtaining labeled private data from users. As the sensitive side-channel metadata information is not used in applications such as activity recognition~\cite{liu2019privacy, boutet2021dysan}, we propose to indiscriminately perturb metadata information while retaining raw activity features. This is achieved by reconstructing raw data from a carefully-designed diffusion-based process, which does not require private attribute labels for training.

\textit{Second, how to dissociate fine-grained activity features from sensing data to prevent sensitive activity information leakage?} Existing approaches such as adding noise or blocking sensitive data could damage the data utility. In many cases such as fall detection, some fine-grained features (e.g., hand gestures) are unnecessary. Yet, discarding sensing data to eliminate fine-grained features would result in the loss of utility. To dissociate sensitive activity information, \S removes fine-grained features by reconstructing multi-grained data. Considering the hierarchical nature of human activities, where an activity is composed of a series of atomic actions. We design a latent diffusion guidance model to generate multi-grained data that naturally removes different granularity of activity information. Many apps such as travel apps and pedometer apps rely on data in its raw form rather than latent representations or activity labels, necessitating the process of data generation.

We evaluate the information leakage in multi-grained sensor data generated by \S using different attribute inference and activity recognition tasks. For metadata-level protection, \S can reduce the private attribute inference probability to 50\%, while maintaining the same level of activity recognition accuracy as raw data. For feature-level protection, \S reconstructs multi-grained data with different levels of activity recognition performance. We also showcase \S on the pedometer application to demonstrate the feasibility that users can control data granularity to prevent unintended information leakage without affecting data utility.

The main contributions of this paper are threefold:
\begin{itemize}

    \item We propose a novel system based on the noise diffusion process to anonymize private sensing data while retaining activity recognition performance, without requiring users to specify private data and labels.

    \item We design the hierarchical latent feature guidance diffusion model for multi-grained data generation to achieve fine-grained control of information disclosure and mitigate feature-level overprivileged issues.

    \item We extensively evaluate the information and utility of reconstructed data from \S under metadata-level and feature-level information protection constraints. 
    
\end{itemize}

\section{Related Work}
\label{sec:related}
\subsection{Metadata Information Protection}
There are two main approaches for metadata information protection for data anonymization. First, the adversarial training-based methods~\cite{liu2019privacy, boutet2021dysan} utilize a discriminator model to infer sensitive attributes from feature representations and a generative model to minimize the success rate of the discriminator. Second, variational autoencoders and information theory-based methods~\cite{zheng2022infocensor} aim to minimize the mutual information between private attributes and their representations. 

However, most methods require users to provide their private data and labels for model training. This is because the design of loss functions in the model optimization needs labeled private attributes. In contrast, \S's model training does not need any private information such as private attribute labels or sensitive activity patterns. 
Moreover, existing noise-based methods not only destroy metadata information but also degrade the utility of activity recognition~\cite{malekzadeh2018replacement}. By contrast, \S perturbs the private attributes and preserves much of the activity information in the raw data for downstream tasks.

\subsection{Semantic Information Protection}
If the fine-grained activity semantic is sensitive, a naive idea is to add adaptive noise to degrade activity recognition performance. However, it is hard to balance the added noise and data utility. Differential privacy adds noise to cover the existence of any data record~\cite{privtrace23wang}, but it targets a different problem than the feature-level overprivileged issue. Therefore, we propose to dissociate fine-grained motion features and only retain coarse-grained activity information in released data. Thus, \S generates multi-grained data, offering a more controllable way for the balance between data utility and privacy protection.

Another idea is to generalize labels from activity recognition APIs~\cite{google2022core}. Yet, obtaining activity labels from trusted OS API remains challenging due to the variations in human behaviors and diverse hardware. More importantly, merely generalizing activity labels does not necessarily provide a privacy guarantee. Thus, many studies~\cite{gotz2012maskit, chakraborty2014ipshield} offer privacy-checking methods for deciding whether to release or suppress activity semantics. If an activity has privacy implications, the existing methods brutally discard the entire segment regarding the activity. However, many apps require the raw data. For instance, a game app utilizes motion sensor data to interpret user gestures, and a travel app relies on geomagnetic and accelerometer sensors to provide a compass bearing. Thus, for apps demanding raw sensor data, existing semantic privacy protection mechanisms are too coarse-grained. By contrast, for semantic-level protection, instead of blocking the whole sensitive data, \S dissociates sensitive features and releases non-sensitive but useful data to apps. In addition, \S can be integrated with existing privacy checking mechanisms~\cite{gotz2012maskit, chakraborty2014ipshield} to achieve fine-grained information disclosure with multi-grained data generation.

\section{Problem Formulation}
\label{sec:problem}
\subsection{User Privacy Leakage in Activity Sensing Data}
\label{subsec:threat}
User privacy and system security are widely studied in literature~\cite{wang2023vsmask, chen2023understanding, guo2023phantomsound, wang2023patch, li2023echoattack}. In this work, we consider personal attributes and activity features that are unintended to be released as private information in the context of mobile activity sensing. Such privacy leakage can be achieved via different attacks such as \textit{attribute inference attacks} and \textit{activity inference attacks} in diverse activity recognition scenarios. The sensing data can be leaked to malicious third parties by data trading or stealing. First, in attribute inference attacks, an adversary aims to infer personal attributes (\eg, age, gender)~\cite{iwasawa2017privacy}. For example, suppose a user adopts behavioral biometrics such as hand gesture~\cite{cheon2020gesture} for authentication. The attackers can extract a large amount of personal attribute information such as gender and age from behavioral biometrics data. Second, in activity inference attacks, an adversary can obtain fine-grained activity semantics from mobile sensing data. For instance, an attacker can utilize the pedometer app to obtain sensitive daily schedules of users by implementing activity recognition using collected sensor data. Moreover, an attacker can steal passwords by modeling hand movements with motion-sensing data~\cite{shukla2019stealing}.

Many activity recognition models can achieve more than 95\% accuracy~\cite{chen2021deep}, which have become a powerful tool for divulging private activity information. In addition, we consider a strong attacker who can implement the re-identification attack~\cite{hajihassnai2021obscurenet}. Specifically, the attacker knows the defense methods and has a dataset with private attribute labels. Then, they can build a sanitized version of the dataset by passing the dataset through the defense methods. Finally, the attacker can train a new private attribute identification model based on the sanitized datasets. To protect activity-sensing information, we assume attackers can only access sensor data after being processed by \S. We assume the mobile sensors are trustworthy as they reside in the secure processing domain (e.g., ARM TrustZone~\cite{arm_trust_zone}). \S can become a middleware between OS system and various applications.

\subsection{Mobile Activity Sensing Data Utility}
\label{subsec:motion}


For sensing data $X$ such as the IMU sensor data, the activity semantic label of $X$ and recognition confidence score are obtained by activity recognition models~\cite{wang2018socialite, chen2021deep}. If fine-grained activity features in $X$ are eliminated, $X$ can only be used for recognizing coarse-grained activity semantics. Our key insight is that human activities naturally follow a hierarchical semantic structure, which plays a crucial role in defining and modeling multi-grained activities~\cite{liu2016lasagna}. An activity can be decomposed into atomic actions or generalized into coarse-grained activity semantics. For example, \textit{Jumping} and \textit{running} can be generalized into \textit{locomotion}. Taking a deeper look into \textit{jumping}, it includes atomic actions such as \textit{arms swinging} and \textit{legs contracting}. 

We define activity sensing data utility function as $U(a, \hat{a}) = \frac{U_{\hat{a}}}{U_a}$ under unintended sensing information disclosure constraints. $U_a$ is the recognition confidence score on raw data $a$ and $U_{\hat{a}}$ is the confidence score on reconstructed data $\hat{a}$. $U(a, \hat{a})$ quantifies the utility of data after generalizing activity semantics from $a$ to $\hat{a}$. For example, raw sensor data contains the highest-resolution features corresponding to fine-grained activity semantics. The maximum utility is 1 if we use the raw data. $U(\cdot)$ monotonically decreases when the semantic of $X$ is generalized from a fine-grained $\hat{a}_{i-1}$ to a coarse-grained $\hat{a}_i$:
\begin{equation}
   0< U(a, \hat{a}_i)\leq U(a, \hat{a}_{i-1}),
\end{equation}
where $i$ is related to the number of stacked convolutional autoencoders (CAEs) in Section~\ref{sub:prior}. Users can designate the output features of the $i$-th stacked CAEs for the $i$-th granularity of data generation, noted as \textit{Granu.$i$}, which balance the privacy and utility needs. 

A third party auditor can help users evaluate the utility of different granularity of reconstructed data. 
For example, a fall detection application on a smartwatch can collect the raw sensing sequence data $X_{hand}$ that represents ``hand gesture". $X_{hand}$ contains high-resolution features that can recognize ``hand position". A user can set $i=0$, and then \S generates \textit{Granu.0} data by only perturbing the metadata information while keeping original features in raw data. Thus, the \textit{Granu.0} data can still be used for fall detection while discarding sensitive metadata information. If $i=1$,  \S generates \textit{Granu.1} data with the guidance of the output feature from the first stacked CAEs, which removes the high-resolution features including ``hand gesture". Then, the \textit{Granu.1} data $X'_{hand}$ can still be used in fall detection but only be recognized as coarse-grained semantics such as ``walking". Similarly, if features of ``walking" are removed in reconstructed data, it can only be recognized as ``locomotion". In this manner, \S helps users avoid releasing sensitive sensing information.

\begin{figure*}[ht]
\centering
\includegraphics[width=0.95\linewidth]{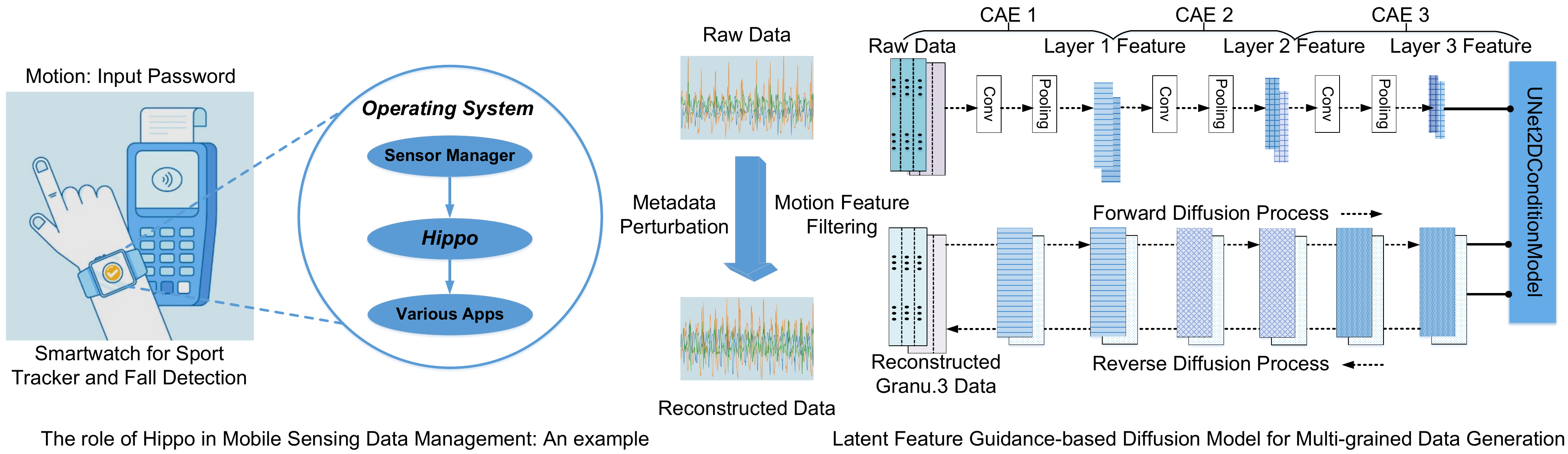}
\caption{The latent feature guidance diffusion model of \S. \S acts as a middleware between the OS Sensor Manager and Apps. \S reconstructs raw data and the Apps obtain data from \S. In multi-grained data generation, right arrows are the forward diffusion process, and left arrows refer to the reverse diffusion process conditioned on multi-resolution layer features.}
\label{fig:system}
\vspace{-15pt}
\end{figure*}

\section{System Design}
\label{sec:system}
\subsection{System Model}
\S runs as a middleware between the OS sensor manager and various apps as shown in Fig.~\ref{fig:system}. The essential component of \S is the latent feature guidance-based diffusion model which carries out hierarchical information dissociation. During the training phase, first, we propose a self-supervised stacked CAEs (\eg, CAE1) to extract multi-resolution activity features (\eg, Layer 1 Feature) without requiring users' private labels (Section~\ref{sub:prior}). Then, conditioned on multi-resolution features, we design a hybrid conditional and unconditional diffusion model to reconstruct multi-grained data (Section~\ref{sub:data_gen}).  During the generation phase, for instance, as shown in Fig.~\ref{fig:system}, \S reconstructs \textit{Granu.3} data conditioned on \textit{Layer 3 features} from the third CAE layer in stacked CAEs to achieve sensitive activity feature dissociation and metadata perturbation.

On one hand, if metadata information is not intended to be released but raw features are required by apps, \S can generate \textit{Granu.0} data to only perturb metadata. We pass the raw data in the forward diffusion process to add noise. Then, in the reverse diffusion process, \S removes the noise with the learning-based procedure. The residual noise after the learning process of denoising in \S reconstructs the raw data by perturbing the side-channel metadata information. Meanwhile, the raw activity sensing data features are retained. 

On the other hand, if fine-grained activity information is not intended to be released, \S first extracts multi-resolution features from a segment of data $X$ with the stacked CAE model. Then, given a random noise seed, conditioned on multi-resolution features, \S can generate multi-grained data using our designed latent guidance diffusion model. The generated data retains specific-layer feature information while discarding finer-grained features. As a result, the generated data contains different granularities of activity semantics expressed by the multi-resolution feature representations. The users can specify the $i$-th stacked CAEs for multi-resolution feature extraction and multi-grained data reconstruction.

\subsection{Multi-Resolution Feature Extraction}
\label{sub:prior}

We propose to apply label-free stacked CAEs to extract multi-resolution features, which guide the generation of multi-grained data. The convolution module (CM) is widely used to extract features from time series. Regarding the utility of extracted features, the advantages of CM are two-fold. First, CM can exploit local connectivity by focusing on multiple receptive fields of the input, which is suitable for time series data where local patterns are often more informative than individual data points. Second, CM applies weight sharing, meaning the same weight is applied across different filters. Each filter can capture patterns across different dimensions of time series. This ensures that the model can recognize a pattern regardless of its position in the time series. Therefore, CM can effectively extract robust features for activity recognition.

\S packs raw sensor data into a structured tensor to feed into stacked CAEs as shown in Fig.~\ref{fig:system}. For each CAE, the input is raw data $X$ or the output feature $z_i$ of the previous CAE. The output of CAE is the feature $z_{i+1}$, where $i$ is the index of stacked CAEs. For example, the input of CAE-1 is $X$, and the output of CAE-1 is $z_1$, which is the input of CAE-2.
Normally, $z_{i+1}$ is in lower dimensional space than $z_i$. But to keep the same spatial dimension between $z_i$ and $z_{i+1}$, we pad $z_{i+1}$ with zeros to the same dimension of $z_i$. 
The encoder of a CAE contains a convolutional layer and a pooling layer. In the encoding process, the convolutional layer and pooling layer learn multi-resolution activity features. The decoder of a CAE contains an unpooling layer and a deconvolution layer. In the decoding process, the unpooling and deconvolution layers are used to reconstruct the input data in each CAE. We train each CAE separately. The objective is to minimize the mean square error between input data and reconstructed data. After the learning process, we can use the encoder of stacked CAEs to extract multi-resolution activity features.

\begin{lem}
Given two stacked  CAEs whose outputs are layer features $z_i$ and $z_j$ respectively. Then $z_j$ is low-resolution version of $z_i$ such that $\mathcal{H}(z_j)<\mathcal{H}(z_i)$ when $i<j$.
\label{lemma:info}
\end{lem}

Different resolutions of feature representations express varied amounts of activity semantic information, composing a hierarchical information structure. Existing studies in visual analytics demonstrated that the convolutional layer can learn different aspects of data such as detecting edges and lines, and the deep convolutional neural network can capture multi-grained feature representations. For instance, the raw feature $z_0$ is regarded as the semantic granularity-0, which contains the most fine-grained motion features in \textit{Granu.0} data. The layer-1 feature $z_1$ corresponds to the semantic granularity-1, which contains coarse-grained motion information by blurring features in $z_0$ in \textit{Granu.1} data. The convolutional and pooling operations cause $z_1$ to dissociate detailed features in $z_0$. The information on the upper layer features $z_i$ will decrease as the number of stacked CAEs increases. 

We use the H-score $\mathcal{H}$ proposed in~\cite{xu2022information} to quantify the informativeness of extracted features. $\mathcal{H}$ is computed in an information-theoretic framework as follows:
\begin{equation}
    \mathcal{H}(s) = \frac{1}{2} \mathbf{E}_{P_Y}[||\mathbf{E}_{P_{X|Y}}[\wedge_{\widetilde{s}}^{-1/2}\widetilde{s}(X)|Y]||^2].
\end{equation}
The input of the $\mathcal{H}$ function is the feature embedding matrix and label matrix of data samples. $\mathcal{H}$ measures the quality of features generated at any layer of the DNN. The larger value of $\mathcal{H}$ means more information in the feature embeddings is related to the label. We compute the H-score on the feature embeddings generated from stacked CAEs using the DSA dataset~\cite{barshan2014recognizing}. Layer-1 feature H-score, layer-2 feature H-score, and layer-3 feature H-score are 7.717, 7.518 and 7.423, respectively. Therefore, using the framework, we show that stacked CAEs can extract multi-resolution features that contain different amounts of information. In the next section, we design the data generation model conditioned on the $z_i$, which generates data containing multi-grained features.

\subsection{Multi-grained Sensing Data Generation}
\label{sub:data_gen}
Suppose an app requires the $i$-th granularity of sensing data, which corresponds to the $i$-th layer feature $z_i$ in the stacked CAEs. With the well-trained latent feature guidance diffusion model, \S generates the $i$-th granularity data from random noise conditioned on a specific resolution feature $z_i$. Specifically, in the offline training stage, the raw data $x$ can be encoded into a bipartite latent representation $(x_T, z_i)$. The diffusion latent $x_T$ is generated from the forward diffusion process, which is the noisy version of $x$ after many steps of adding noise. $x_T$ encodes the residual information necessary for reconstructing $x$ under the reverse diffusion process. The multi-resolution feature $z_i$ is generated by the stacked CAE, and $z_i$ describes the multi-grained feature representations. 

Following notations in the work~\cite{ho2020denoising} about denoising diffusion probabilistic models, in the training phase, the diffusion model (DM) adds random noise $\epsilon_t\sim\mathcal{N}(0, \sigma_t^2)$ to input data with a noise scheduler. Then, DM learns to remove $\epsilon_\theta(x_t, t)$, which is the predicted noise. The general learning objective is 
\begin{equation}
\begin{split}
    L &= ||\mu_t-\mu_\theta|| = \mathcal{E}_{t\sim[1,T], x_0, \epsilon_t}[||\epsilon_t-\epsilon_\theta(x_t, t)||^2] \\
    &=\mathcal{E}_{t\sim[1,T], x_0, \epsilon_t}[||\epsilon_t-\epsilon_\theta(\sqrt{\bar{\alpha}_t}x_0+\sqrt{1-\bar{\alpha}_t}\epsilon_t, t)||^2],
\end{split}
\label{eq:loss}
\end{equation}
where $\mu_t$ is the posterior mean in the forward diffusion process, and $\mu_\theta$ is the predicted mean in the reverse diffusion process. $\sqrt{\bar{\alpha}_t}$ is related to the noise schedule in the forward $T$ diffusion steps, and $x_0$ is the raw input data. 

A key observation is that the learning objective $L$ in Eq.~(\ref{eq:loss}) relies on the marginal $q(x_t|x_0)$ instead of the joint $q(x_{1:T}|x_0)$. The forward process can be derived from Bayes’ rule~\cite{song2020denoising}:
\begin{equation}
    q_\sigma(x_t|x_{t-1},x_0) = \frac{q_\sigma(x_{t-1}|x_t, x_0)q_\sigma(x_t|x_0)}{q_\sigma(x_{t-1}|x_0)}.
\end{equation}
Therefore, each $x_t$ depends on $x_{t-1}$ and $x_0$, making the process a non-Markovian process. Thus, $L$ in Eq.~(\ref{eq:loss}) does not rely on a particular forward procedure. Therefore, we can sample part of the forward steps during the generation process if $q(x_t|x_0)$ is fixed. In this way, we can reduce the number of steps $T$ and improve the data generation efficiency. 

In the reverse diffusion process, \S learns to remove noise in $X_T$ that is added in the forward diffusion process. Specifically, we design \S to reconstruct back to the original data $x_0$ from $x_T$ conditioned on the feature $z_i$. The guidance from $z_i$ helps the reconstructed data contain the same feature representation information as $z_i$ in the latent space. Consider the generative model is formalized as $P(x|z)=\frac{P(x,z)}{P(z)}$, and $P(z)$ is a deterministic function of $x$ in the multi-resolution feature extraction model, we only need to estimate $P(x,z)$. Thus, we propose to project the $z_i$ to the existing diffusion latent $x_T$: 
\begin{equation}
    \epsilon_\theta(x_T|z_i, T)=\epsilon_\theta(x_T|\emptyset, T)+s\cdot(\epsilon_\theta(x_T|z_i, T)-\epsilon_\theta(x_T|\emptyset, T)),
    \label{eq:z_guide}
\end{equation}
where $s$ is the guidance scale, which is empirically set as 7.5 in our experiments. In Eq.~(\ref{eq:z_guide}), we can simultaneously train an unconditional diffusion model $p_\theta(x_T)$ and a conditional diffusion model $p_\theta(x_T|z_i)$. For the unconditional model, we replace the $z_i$ with a null vector with a probability such as 0.3. For the conditional model, the denoising process is conditioned on $z_i$. Thus, we can guide the data generation towards $z_i$ during the training phase.

To guarantee $q(x_t|x_{t-1})$ still satisfies conditional distribution that follows the Markov chain in the forward process, the reverse process $q_\sigma$ should satisfy the Gaussian function~\cite{song2020denoising}:
\begin{equation}
    q_\sigma (x_{t-1}|x_t, x_0)=\mathcal{N}(\mu, \sigma_t^2I),
\end{equation}
\begin{equation}
x_0=\frac{x_t-\sqrt{1-\alpha_t}\epsilon_\theta(x_t|z_i, t)}{\sqrt{\alpha_t}},
\end{equation}
\begin{equation}
    \mu = \sqrt{\alpha_{t-1}}x_0+\sqrt{1-\alpha_{t-1}-\sigma_t^2}\cdot\epsilon_\theta(x_t|z_i, t).
    \label{eq:non_mu}
\end{equation}
Thus, given a subset $\{x_{\tau_1}, \cdots, x_{x_{\tau_S}} \}$ where $\tau$ is a sub-sequence of $[1,\cdots, t, \cdots, T]$ of length $S$, the model can be trained with arbitrary forward steps. We follow the setup to variance $\sigma$ in~\cite{song2020denoising} as follows:
\begin{equation}
    \sigma_{\tau_i} = \sqrt{(1-\alpha_{\tau_{i-1}})(1-\alpha_{\tau_{i}})}\sqrt{1-\alpha_{\tau_{i}}/\alpha_{\tau_{i-1}}}.
\end{equation}

It has been proven that the training objective in Eq.~(\ref{eq:loss}) is still applicable in the reverse process~\cite{song2020denoising}. Finally, we train the diffusion model to approximate the conditioned probability distribution in the reverse process.
Given the initial input noise $\epsilon_t\sim\mathcal{N}(0,I)$, we sample $x_{t-1}\sim p_\theta(x_{t-1}|x_t)$ as follows:
\begin{equation}
\begin{split}
   x_{t-1} &= \mu+\sigma_t\epsilon_t,
   \label{eq:x_t-1}
\end{split}
\end{equation}
where $\mu$ is defined in Eq.~(\ref{eq:non_mu}). With the same model for predicted noise $\epsilon_\theta$, we can choose different $\sigma_t$ without retraining the model to generate different samples. In the sampling phase, the reverse process in the DM can generate multiple data corresponding to a given latent feature representation $z_i$. Particularly, when $\sigma_t\neq 0$, the reverse process is non-deterministic. Larger values of $\sigma_t$ introduce higher stochasticity, resulting in more variations of generated data guided by the same latent representation $z_i$.

\begin{table*}[t]
\centering
\caption{Activity recognition utility and gender attack success rate on anonymized MotionSense dataset.}
\label{tab:case1}
\scalebox{1.1}{
\begin{tabular}{lcccccccccc}
\toprule
\multicolumn{1}{c}{\multirow{3}{*}{\textbf{Method}}} & \multicolumn{8}{c}{Average Classification Accuracy on 5-fold Cross-Validation (\%)}                                                                               & \multicolumn{2}{c}{\multirow{2}{*}{Acuracy Overall}} \\ \cline{2-9}
\multicolumn{1}{c}{}                                 & \multicolumn{2}{c}{Downstairs} & \multicolumn{2}{c}{Upstairs} & \multicolumn{2}{c}{Walking} & \multicolumn{2}{c}{Jogging} & \multicolumn{2}{c}{}                                 \\ \cline{2-11} 
\multicolumn{1}{c}{}                                 & Activity        & Gender       & Activity       & Gender      & Activity      & Gender      & Activity      & Gender      & Activity                   & Gender                  \\
\midrule
Raw Data                                        & \textbf{95.6}           & 87.7        & \textbf{95.4}          & 90.9       & \textbf{98.5}         & 95.1       & \textbf{97.3}         & 95.6       & \textbf{96.7}                      & 92.3                   \\
Differential Privacy~\cite{dwork2016concentrated}                                 & 82.5           & 78.3        & 83.3          & 79.1       & 84.6         & 80.2       & 84.2         & 79.8       & 83.7                      & 79.4                   \\
AAE~\cite{malekzadeh2019mobile}                                                   & 90.7           & 59.8        & 93.2          & 57.6       & 91.8         & 52.9       & 93.7         & 52.5       & 92.4                      & 55.7            \\
TIPRDC~\cite{li2020tiprdc}                                               & 91.5           & 77.4        & 92.8          & 78.6       & 97.5         & 79.4       & 95.2         & 78.8       & 94.2                      & 78.6                   \\
ObscureNet~\cite{hajihassnai2021obscurenet}                                           & 87.5           & 72.5       & 92.8          & 81.3        & 94.7         & 83.2       & 96.9         & 89.5       & 92.9                      & 81.6                    \\
InfoCensor~\cite{zheng2022infocensor}                                           & 91.1           & 76.8        & 92.4          & 77.1       & 96.6         & 78.2       & 94.3         & 77.3       & 93.6                      & 77.4                   \\
\textbf{Hippo-Granu.0}                                                & 92.4           & 54.2       & 92.8          & 54.8       & 97.6         & \textbf{51.6}       & 96.5         & \textbf{50.7}       & 94.8                      & 52.8               \\
Hippo-Granu.1                                                & 83.8           & 52.1        & 84.2          & 53.5       & 86.3         & 53.5       & 84.2         & 51.8       & 84.6                      & 52.7               \\
Hippo-Granu.2                                                & 66.2           & 53.5        & 65.7          & 52.8       & 63.8         & 54.7       & 67.2         & 51.8       & 65.7                      & 53.2               \\
Hippo-Granu.3                                                & 53.8           & \textbf{51.8}        & 55.4          & \textbf{52.1}       & 54.7         & 53.5       & 53.6         & 52.5       & 54.4                      & \textbf{52.5}               \\
\bottomrule
\end{tabular}
}
\end{table*}

\section{Implementation and Evaluation}
\label{sec:evaluation}
\subsection{Experimental Setup}
We evaluate \S on four datasets: HARBox~\cite{ouyang2021clusterfl}, UCI HAR~\cite{anguita2013public}, MotionSense~\cite{Malekzadeh:2019:MSD:3302505.3310068}, and the daily and sports activity (DSA) dataset~\cite{barshan2014recognizing}.
For the diffusion model training, we set random timesteps from 100 to 1000 because the model can be trained with any number of forward steps as designed in Section~\ref{sub:data_gen}, and the training epochs are 100. In the multi-grained data generation phase, we set the inference time steps as 100 following the empirical experiments. The guidance scale of hierarchical features is set to 2. For sensitive attribute and activity recognition, we use the 5-fold cross-domain validation considering the training and testing dataset size, so as to avoid overfitting and provide more consistent and fair evaluations.

\subsection{Metadata-level Information Evaluation}
\label{sub:case1}

We evaluate the ability of \S to defend against metadata inference attacks while retaining raw activity features. We consider a scenario where users do not want to release their personal metadata information and applications require raw activity features. The adversary aims to obtain metadata information behind the released sensing data following the threat model in Section~\ref{subsec:threat}. 
In this case, \S generates Granu.0 data by only perturbing the attribute information while keeping the fine-grained activity features in raw data. 

We compare \S with five baseline methods: (i) DP~\cite{dwork2016concentrated} method injects Laplace noise with various privacy budgets into the sensing data. We report the average accuracy under four privacy budget parameters $\{0.1, 0.3, 0.7, 0.9\}$ to reflect the general balance between utility and privacy. (ii) Anonymization Autoencoder (AAE)~\cite{malekzadeh2019mobile} is a deep autoencoder model to minimize the identification information and preserve the original activity information. (iii) TIPRDC~\cite{li2020tiprdc} hides private information by training a mutual information estimator and the sanitized features are used for activity recognition. (iv) ObscureNet~\cite{hajihassnai2021obscurenet} is an encoder-decoder structure to conceal private attributes in time-series data. The synthesized data is used for activity recognition. (v) InfoCensor~\cite{zheng2022infocensor} is an information-theoretic framework to minimize the mutual information between the representations and the attributes of raw data.

The performance of different models for attribute protection is shown in Table~\ref{tab:case1}. Taking the gender attribute as an example, these models are trained to sanitize the gender attribute. We use two convolutional neural network models on the MotionSense dataset with 5-fold cross-validation to perform gender and activity recognition. For noise-based DP, though we can add stronger noise to preserve sensitive information better, it largely destroys the data utility. For instance, when the added noise in DP degrades the gender recognition accuracy from 92.3\% to 79.4\%, the activity recognition accuracy also drops from 96.7\% to 83.7\%.
By contrast, \S allows the sensitive attributes in the reconstructed data \textit{Granu.0} to be eliminated while retaining as much of the original activity semantic information as possible. For example, the gender recognition accuracy on the \textit{Granu.0} data is only around 52.8\%, reaching the level of a random guess. Meanwhile, the activity recognition accuracy on \textit{Granu.0} data is around 94.8\%, which is comparable to model performance on raw data. \S retains high data utility because of the learning-based denoising process in the reverse diffusion.

A number of methods such as AAE, TIPRDC, ObscureNet, and InfoCensor are based on information minimization in adversarial training. For example, ObscureNet modifies the private attributes in the latent features before synthesizing a new version of data. Thus, it can fool the gender inference model to an accuracy of around 20\%. However, for a binary classification task, reversing the output is trivial for a strong attacker who knows the mechanism of ObscureNet. Thus, we report the reversed accuracy in Table~\ref{tab:case1}. For example, for the ObscureNet method, the gender inference accuracy can be reversed, achieving around 80\% accuracy. In contrast, \S decreases the gender inference probability to around 50\%, which causes the best entropy (uncertainty) for attackers. Besides, we also evaluate the gender information in multi-grained data. The multi-grained data generation is based on the random noise conditioned on multi-resolution feature representations. As a result, from Table~\ref{tab:case1}, we can see the gender information is also perturbed in the multi-grained data, rendering gender recognition a random guess. In summary, for metadata-level overprivileged information protection, \S effectively preserves data utility while perturbing overprivileged attributes.

\subsection{Feature-level Information Evaluation}
In this section, we consider users have fine-grained sensitive motion features that are unintended to be released. 

\begin{figure}[t]
\centering
\includegraphics[width=\linewidth]{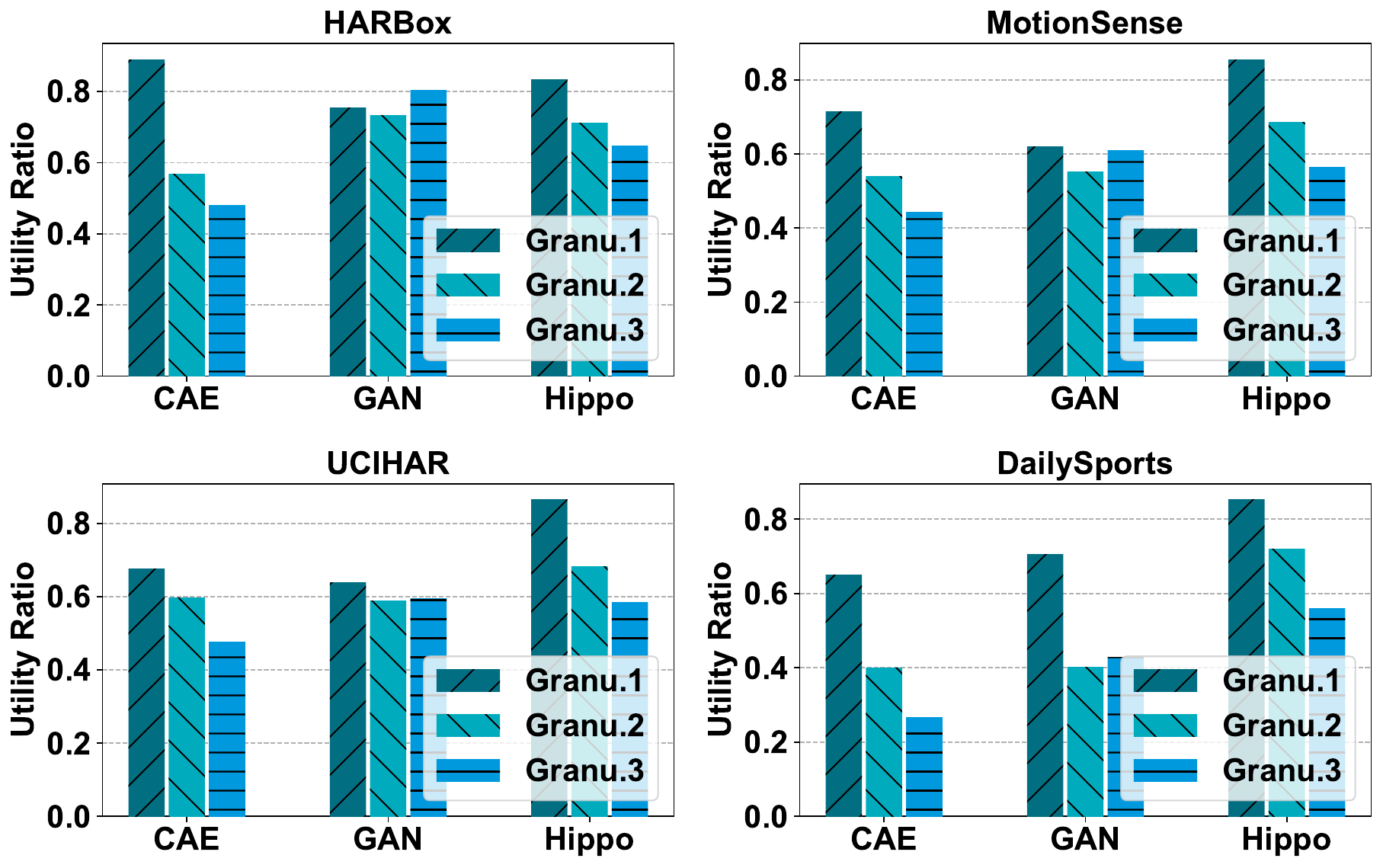}
\caption{Comparison of different generative models.}
\label{fig:bar_plot}
\end{figure}

\subsubsection{Classification}
We compare \S with two other generative models: Convolutional Autoencoder (CAE)~\cite{vincent2008extracting} and Generative Adversarial Network (GAN)~\cite{goodfellow2020generative}. (i) For CAE, we extract the $i$-th layer feature $z_i$ of raw data $X$ with the encoder of CAE. Then, we map the $z_i$ with a matrix $W$ to the data $X'$: $X'=Wz_i$. We control the error rate between $X$ and $X'$: $Err = ||X'-X||_2$ to retain the amount of information in $X'$. For CAE data evaluation, we train the CNN model on datasets generated by CAE. For CAE, we set three thresholds \{0.001, 0.01, 0.1\} to generate three levels of granularity of data.

(ii) GAN is based on adversarial training to learn the generation of new data. We design GAN to implement the adversarial information factorization~\cite{creswell2017adversarial} based on the specified layer of latent representations. For GAN, we use adversarial training to train the generator conditioned on layer 1 to 3 features. We use the same CNN model structure (two convolution layers and two fully connected layers), but the CNN models are trained on different datasets generated by different generative models. We use \S to generate new data with the guidance of layer 1 to 3 features. Fig.~\ref{fig:bar_plot} shows the differences in the utility on the different granularity of data.

\S generates multi-grained data with different amounts of semantic information. For instance, on the MotionSense dataset, \S achieves the best utility score of 0.856 for the Granu.1 reconstructed data. The Granu.1 data guided by the layer-1 latent features still preserves the most activity information of raw data. The utility scores in the Granu.2, and Granu.3 data are 0.685 and 0.564, respectively. The decrease in the utility score means the accuracy of recognizing fine-grained activity semantics on coarse-grained data (i.e., Granu.2 and Granu.3) decreases. Thus, the results show that  \S can generate multi-grained data containing different resolutions of activity features. \S dissociates the fine-grained activity features during the data reconstruction process, removing the fine-grained activity semantics from the data.

\S outperforms the existing generative models in the following aspects. First, although the CAE method can also control the information in the generated data, it is hard to set the $E_{rr}$ to control the data granularity because of the heterogeneity of sensor data. For example, different sensors or activities will have different magnitudes of data. For CAE, the utility scores of Granu.1 data on HARBox, MotionSense, UCIHAR, and DailySports are 0.89, 0.715, 0.676, and 0.65, respectively. In contrast, for \S, the utility scores of Granu.1 data on four datasets are 0.834, 0.856, 0.864, and 0.852. The utility scores of CAE on the same granularity show higher divergence than scores of \S. 

Second, for the GAN method, we find that it is hard to train GAN to generate high-quality data steadily in practice. Though GAN is designed to be conditioned on latent features, it is difficult to control the amount of information contained in the generated data. For example, a CNN model trained on the Granu.2 data has better performance than the model trained on the Granu.1 data. Moreover, the data generated by GAN exhibits more noise, resulting in a more diverse utility compared with other methods as shown in Fig.~\ref{fig:bar_plot}. However, \S can generate high-quality data steadily, indicated by the stable utility across different datasets.

\subsection{Ablation Study and Efficiency Analysis}
\subsubsection{The impact of input feature}
\begin{figure}[t]
\centering
\includegraphics[width=\linewidth]{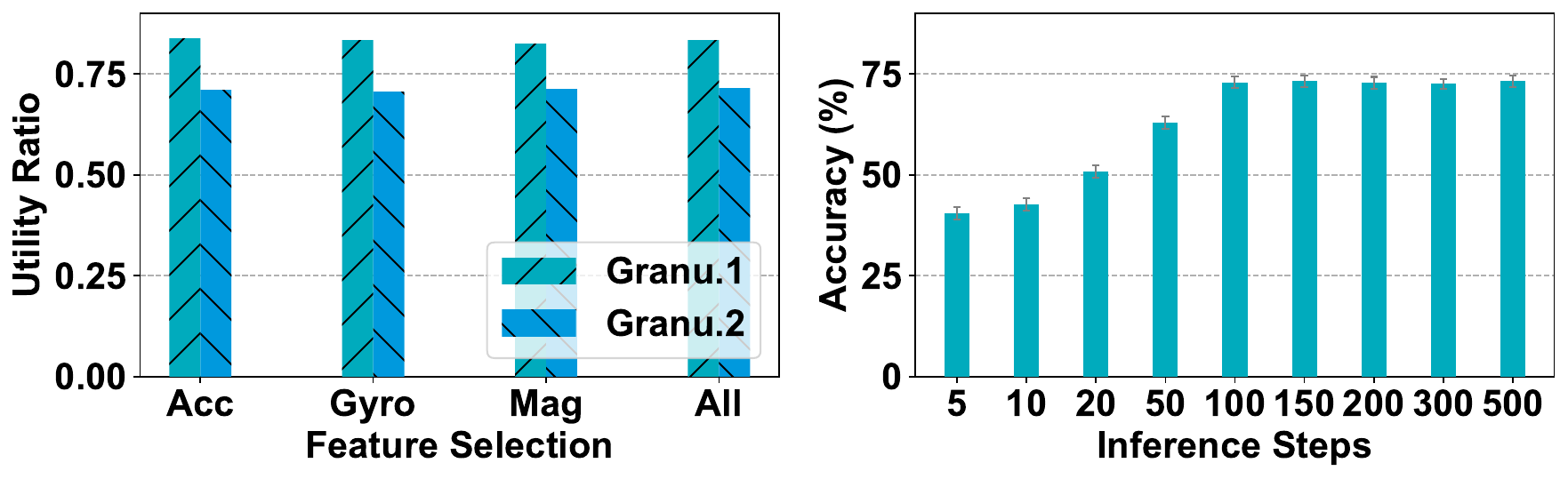}
\caption{The impact of modality and inference steps. Acc. means 3-axis in accelerometer modality, Gyro. means gyroscope, Mag. means magnetometer, and All means the 9-axis modalities. The utility ratio is the accuracy of the model on reconstructed data over the accuracy of raw data.}
\label{fig:ablation}
\end{figure}

We first evaluate the impact of data modality on the ability of multi-grained data generation of \S. We train \S on the HARBox Dataset with different modalities. Then, we generate different granularity of data with the multi-grained latent representations of different modalities. We report the activity recognition utility of reconstructed data in Fig.~\ref{fig:ablation}. For instance, for Granu.1 reconstructed data, the utility ratios on the four modalities are 0.836, 0.833, 0.823, and 0.834, respectively. The utility ratios are similar across different selections of features, which corroborates the robustness of \S on the selection of different modalities. 

\subsubsection{The impact of inference steps} We evaluate the impact of inference steps in the diffusion process for data generation using the HARBox Dataset. We report the activity recognition accuracy on the Granu.1 data generated by \S in different steps. As shown in Fig.~\ref{fig:ablation}, when the inference step is small such as 5, the recognition accuracy is as low as 40.5\%, which shows that the quality of generated data is low. The accuracy improves as the number of inference step increases to 100. Nevertheless, the accuracy becomes relatively stable after the inference step is larger than 100. The result demonstrates that the large inference step does not necessarily guarantee the quality of data.

\subsubsection{Runtime Efficiency}
Suppose the sensing application sampling rate is 50Hz, and we buffer a 1s time window for data reconstruction. On a server with NVIDIA RTX A6000 graphics cards, for the sequence data with the size $100\times 9$, \S takes only 0.61s to reconstruct the new version of the data with the 100 inference steps in DM. Thus, within the 1s buffer, \S is ready to generate the new data. Moreover, we measure the model size of \S by saving it to a disk file, which is only 49.5 MB. GPU-empowered mobile devices such as the iPhone 15pro can achieve text-to-image generation on mobile phones~\cite{stable-diffusion-coreml-apple-silicon}. Nevertheless, an alternative solution is to run \S on a trusted edge server to process raw sensing data and generate new versions of data, and all the applications could access the mobile sensing data from \S. 

\subsection{Step Counter Case Study}
\label{subsec:case}
To protect the sensitive activity information while keeping the utility of counting steps, the user can use \S to reconstruct the motion sensor data. We collected the IMU sensor data at a 50 Hz sampling rate for a group of activities using two types of smartphones: Lenovo ZUK Z2 and Samsung Galaxy S9. To simplify the analysis, we consider four main activity time windows $X_1, X_2, X_3, X_4$.  $X_1$ is walking, $X_2$ is calling during walking, $X_3$ is jumping, and $X_4$ is waving while holding the phone. Two volunteers walk for 200 steps, 600 steps, and 1000 steps for pedometer evaluation. We report the utility of reconstructed multi-grained data in Table~\ref{tab:2cases}.

To evaluate the generalization ability of \S across different datasets, we use HARBOX and MotionSense datasets to train the generative diffusion model in \S, and then use \S to reconstruct the data in the Pedometer Dataset in different granularity with $i=\{0, 1, 2\}$. We first show the probability of leaking sensitive information ``calling" in the multi-grained data. As shown in Table~\ref{tab:2cases}, the adversary can recognize the calling activity with a high accuracy of 92.8\% on raw data, while only achieving a low accuracy of 78.4\% and 63.8\% on Granu.1 and Granu.2 data, respectively.

\begin{figure}[t]
\centering
\includegraphics[width=\linewidth]{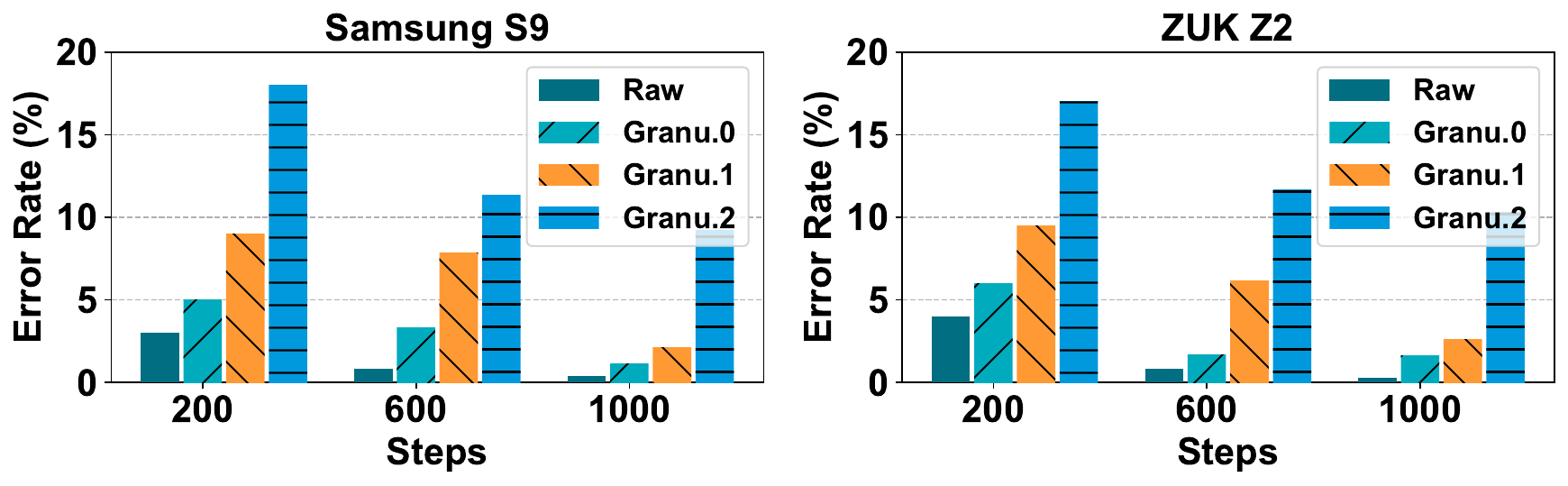}
\caption{The pedometer utility on multi-grained data.}
\label{fig:step_plot}
\end{figure}

\begin{table}[t]
\centering 
\caption{Case study results on multi-grained data.}
\label{tab:2cases}
\scalebox{1.2}{
\begin{tabular}{@{}lcccccc@{}}
\toprule
Granularity & Walking  & Calling & Jumping & Waving         \\
\midrule
Raw Data    & 98.3     & 92.8    & 96.4    & 93.5           \\
Granu.0     & 98.3     & 92.8   & 96.4    & 92.5            \\
Granu.1     & 83.5     & 78.4    & 84.8    & 81.5          \\
Granu.2     & 65.4     & 63.8    & 64.6    & 63.5           \\ \bottomrule
\end{tabular}
}
\end{table}

Then, we evaluate the utility of reconstructed data for step counting using Pydometer~\cite{pydometer2023}. As shown in Fig.~\ref{fig:step_plot}, the Granu.0 and Granu.1 walking data can still preserve the step information as the error rate is below 5\% with the increase of steps to 1000. Therefore, the Granu.1 data still preserves the peak and valley information for the pedometer service. Meanwhile, once we know the average error rate of step counting on Granu.1 data, we can estimate the accurate steps by considering the error rate. Thus, we can set $i=1$ in stacked CAE for data reconstruction to mitigate the overprivileged ``calling" information leakage while keeping the utility of step counting. Nevertheless, suppose the user requires the full utility of the sensor data. In that case, they can set the granularity $i=0$ as Granu.0 to preserve the raw data $X_i$ in a time window while perturbing metadata information. Otherwise, if the user requires more privacy protection, they can set $i=2$, which degrades the calling activity recognition accuracy to 63.8\%. It is at the user's discretion to set different granularity $i$ to prevent unintended information leakage while maintaining the application utility.

\section{Conclusion}
\label{sec:conclusion}
In this work, we explore metadata-level and feature-level overprivileged issues in mobile sensing applications. We design a multi-grained data generation model based on the diffusion model, enabling users to perturb private metadata information and dissociate fine-grained sensitive activity information. \S perturbs the private attributes and eliminates fine-grained activity features by generating data with different levels of features. \S allows users to control the disclosure of the activity information in activity sensing data. \S contributes to the field by empowering users to have fine-grained control over activity information disclosure on mobile devices. 

In the future, we will provide theoretical proof for the quantification of privacy protection. Meanwhile, we acknowledge that data granularity and the hierarchical activity semantics (which depend on the specific applications) may not have a direct one-to-one relationship. With the development of deep learning theory, we will explore the general and explainable relationship between data granularity and activity semantics. Thus, we can provide an interpretable way to determine which granularity of privacy information to disclose.



\clearpage

\bibliographystyle{IEEEtran}
\bibliography{references}

\end{document}